# Teaching Light Polarization by Putting Art and Physics Together

Fabrizio Logiurato

Basic Science Department, Ikiam Regional Amazonian University, Ecuador
Physics Department, Trento University, Italy

*Abstract*—Light Polarization has many technological applications and its discovery was crucial to reveal the transverse nature of the electromagnetic waves. However, despite its fundamental and practical importance, in high school this property of light is often neglected. This is a pity not only for its conceptual relevance, but also because polarization gives the possibility to perform many beautiful experiments with low cost materials. Moreover, the treatment of this matter lends very well to an interdisciplinary approach, between art, biology and technology, which usually makes things more interesting to students. For these reasons we have developed a laboratory on light polarization for high school and undergraduate students. They can see beautiful pictures when birefringent materials are set between two crossed polarizing filters. The colourful images remind them of those ones of abstract painting or alien landscapes. With this multidisciplinary teaching method students are more engaged and participative and also the learning process of the respective physics concepts is more effective.

## I. Introduction

WE have generally found that many students do not like science because, according to them, this does not generate a particular emotional feeling as art or literature. However, according to the philosopher Karl Popper, science is not only like art but it is the most human of all the arts. Keeping in mind Popper's phrase, we have developed a multidisciplinary and artistic approach to the study of light polarization. Light polarization has many important utilizations [1]. It is used in technological devices from chemistry to medicine, from optics to engineering and also in the natural world we can find a lot of its applications. In optical systems, wave plates are made by birefringent materials, liquid crystal displays utilize birefringence and light modulators use electrically induced birefringence. Birefringence is present in many medical devices, for example, it is used to measure the thickness of the optical nerve and the diagnosis of glaucoma. In general, many biological materials are birefringent and microscopy makes use of polarized light in the analysis of tissues. Polarization is also commonly used to study the structure of crystals and molecules. In mechanical engineering birefringence is exploited to detect stress in structures.

Unfortunately, light polarization is usually considered a marginal concept in school programs. In order to fill this gap, we have developed an interdisciplinary laboratory on this subject for high school and undergraduate students. In this one pupils have the possibility to understand the physics of light polarization, its connections with the chemistry and the basics of its applications. Beautiful and fascinating images appear when we set between two crossed polarizing filters some materials. The pictures that we can see are similar to those of the abstract art. Students can create their own artistic compositions, take photos of them and make their experiments with the guide of the teacher. The experimental apparatus is very cheap: they use laptop computer screens as source of polarized light, polarizing filters, water with cane sugar, fructose and other sugars and combinations of glasses with several shapes, many simple objects recycled from the trash, such as plastic boxes or packaging, which are usually birefringent. Indeed, they can create art with garbage.

During their activity, teachers can discuss with pupils the relation between art and science, for instance, how artistic trends in the course of art history were very depending on science and technology. To make just an example, photography or cinema did not kill painting, but they have given new means to communication and expression to human feelings, maybe more hard-hitting than traditional painting.

## II. Playing with Dirac

For our experiments we use polarizing filters of the kind invented by Edwin Herbert Land in 1938. Land was an American scientist, founder of the Polaroid and also inventor of the camera instant photography. Usually modern polarizing filters are made from a sheet of layers with aligned chains of polyvinyl alcohol polymers. The chains are covered with iodine in order to make them conductive. In this way an electromagnetic wave can induce electrical currents along the polymers and these molecules absorb the energy of the wave with polarization along their direction. So the transmitted light is mainly polarized in the direction perpendicular to the polymers (Fig. 1a).

To illustrate the behavior of polarizing filters, students can perform the experiment that Paul A.M. Dirac describes in the introduction of his famous book of quantum mechanics [2].

fabrizio.logiurato@unitn.it, fabrizio.logiurato@ikiam.edu.ec

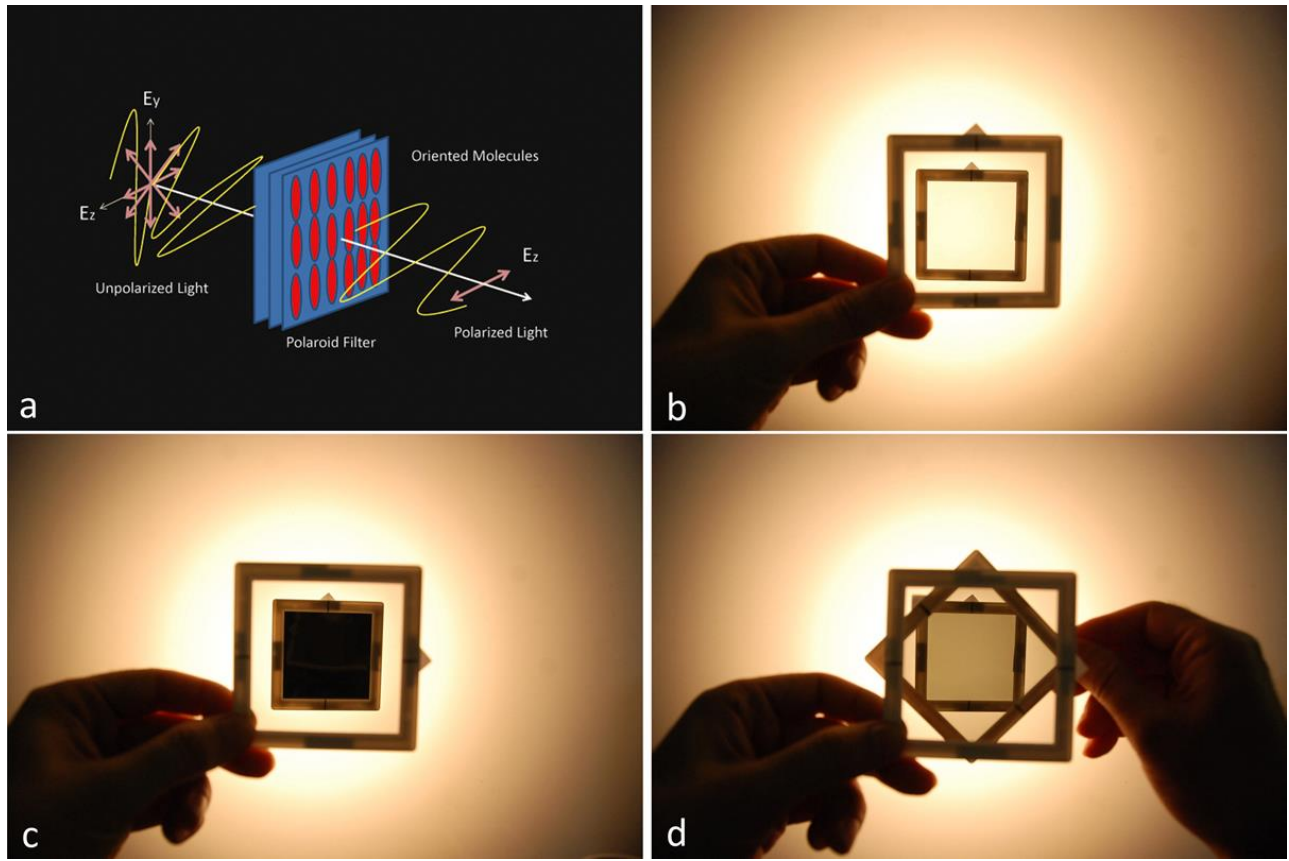

Fig. 1 (a) A polarizing filter is made from aligned polymers which absorb the energy of the wave with vibration along their direction. (b) Unpolarized light emitted by a source is polarized by a Polaroid filter: we can check this by means of a second filter. In (c) the directions of polarization of the filters are orthogonal to each other and the light of the source is almost completely absorbed. But if a third filter is set between the two, as in (d), a fraction of the light can be transmitted again.

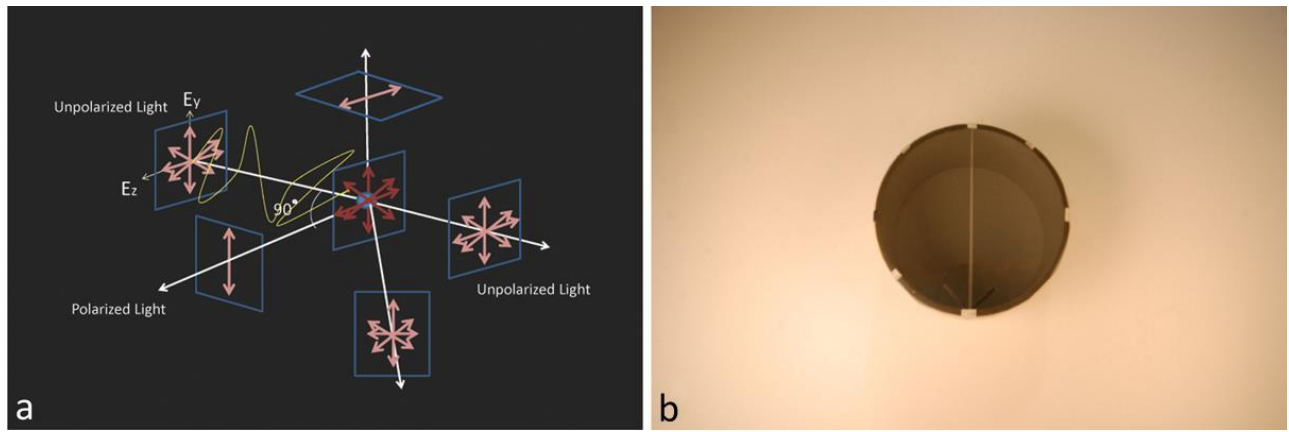

Fig. 2 (a) Unpolarized light becomes partially polarized by scattering. In particular, scattered light in the plane of the dipole oscillations is linearly polarized (b) A Polariscope made with two polarizing filters at right angles to each other. Two little black lines indicate the polarizing directions of the filters.

In Fig. 1b a white screen is illuminated from behind by a lamp that emits unpolarized light. A polarizing filter polarizes the light along a certain direction, as we can note by observing the light through a second filter. If the two filters are perpendicular the light is almost completely blocked (Fig. 1c). But a third intermediate filter can make reappear part of the light, if it creates a component of the light with non-orthogonal polarization to the transmission direction of the last filter (Fig.1d). Pupils are always very impressed by this experiment: it looks like something of magic.

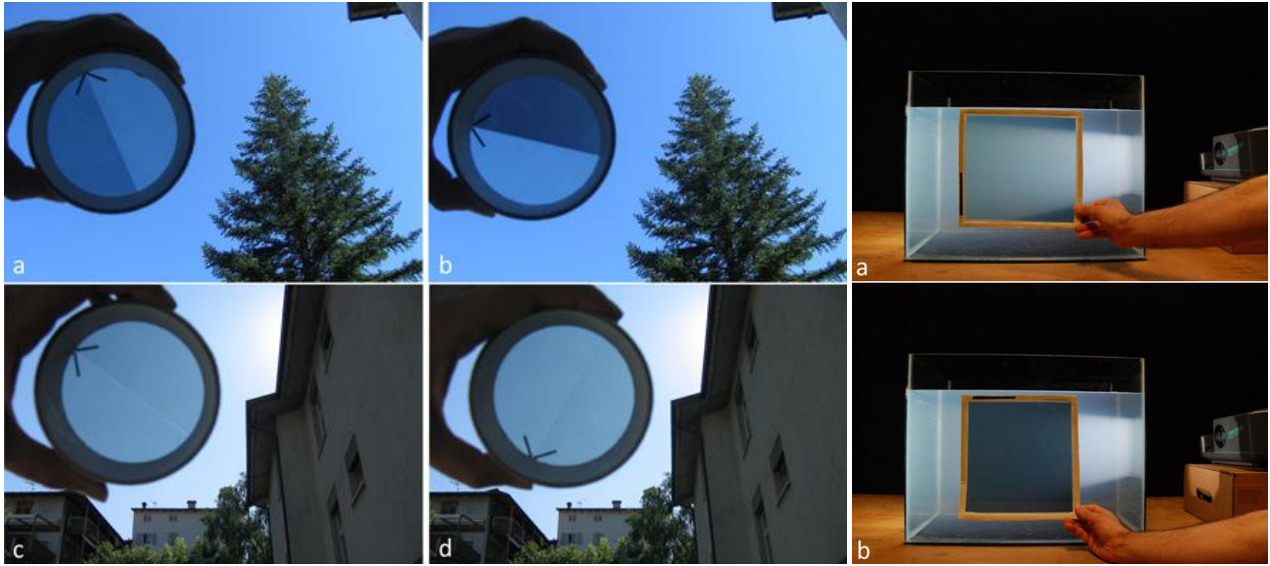

Fig. 3: The polarization of the sky can be determined with a polarizing filter or a polariscope. At 90° degrees with respect to the rays of the sun, light results rather polarized (see (a) and (b) on the left). However, forward the sun, light is very little polarized (on the left (c) and (d)). In the pictures on the right we show a Tyndall experiment on polarization by scattering. In (a) we observe how scattered light has orthogonal polarization with respect to the beam direction. In (b) instead we see how light with parallel polarization with respect to the beam is practically not scattered (the little black line on the filter indicates the polarization of the transmitted light).

## III. KARL VON FRISCH'S DANCING BEES

As observed for the first time by the French physicist Dominique F. J. Arago in 1809, sky light is polarized: the degree of polarization is very high at 90° degrees with respect to the direction of the sunlight and decreases both forwards and backwards [3]. Lord Rayleigh found a theoretical explanation of that [4]. We can give a simple qualitative explanation of Lord Rayleigh's demonstration.

Consider an atom or molecule as an electric dipole (but we can also deal with small particles of smoke or dust, small with respect to the wavelength of the light). When the incident radiation is linearly polarized the electric dipole oscillates along such a direction and the scattered light has its characteristic polarization and intensity (along the direction of oscillation of the dipole there is no scattered radiation). However, even if the incident radiation is unpolarized, the scattered radiation is partially polarized. In fact, the oscillations of the electric dipole induced on the charges are all in a plane perpendicular to the incident radiation. Then the scattered radiation parallel to this plane is polarized (Fig. 2a).

The polarization of light can be better pointed up if we build a polariscope. The simplest polariscope is made by two pieces of polarizing filters with directions of polarization at right angles to each other, as in Fig. 2b. We can use the contrast of brightness between two filters as a more sensitive detector of polarization than a single piece. With a polariscope the sky polarization is made more evident, as we can verify examining the Fig. 3 on the left. According to the explication for the sky polarization by scattering, we could find the direction of the sun rays observing the polariscope in correspondence of its position when we observe a darkest piece of filter. The polarizing direction of such a filter gives us the direction where it is possible to find the sun.

Polarization by scattering can be easily studied in laboratory reproducing Tyndall experiments. John Tyndall in the 1860's showed that small particles like dust, fine smoke or little drops, scatter light of short wavelength more powerfully than long wavelength [5]. This is the reason of the blue color of smoke, for instance. Rayleigh showed that this is true also for air molecules and so he could explain the blue of the sky. Tyndall examined the light scattered by smoke within a long tube of glass and he discovered that it is polarized in the direction normal to the beam. We can reproduce Tyndall experiments using water and a little of milk to increase the scattered light (in this case the molecules of milk fat make the most of the job). In the pictures on the right of Fig. 3 we use an aquarium filled with water and the light beam of a slide projector. We can check that just light with orthogonal polarization with respect to the beam direction is scattered.

Another interesting polariscope was invented by Karl von Frisch. Von Frisch was an Austrian scientist famous for his studies on the physiology of the honey bees, their behavior and the discovered of the meaning of their waggle dance [6].

Waggle dance is called in ethology the odd figure-eight dance that honey bees perform in the hive (Fig. 4a). In this way they share information about the localization of food with their mates. According their language code, the vertical direction indicates the direction of the sun, the angle of the waggle with respect to the vertical shows the direction where the food is with respect to the sun and the hive, while the

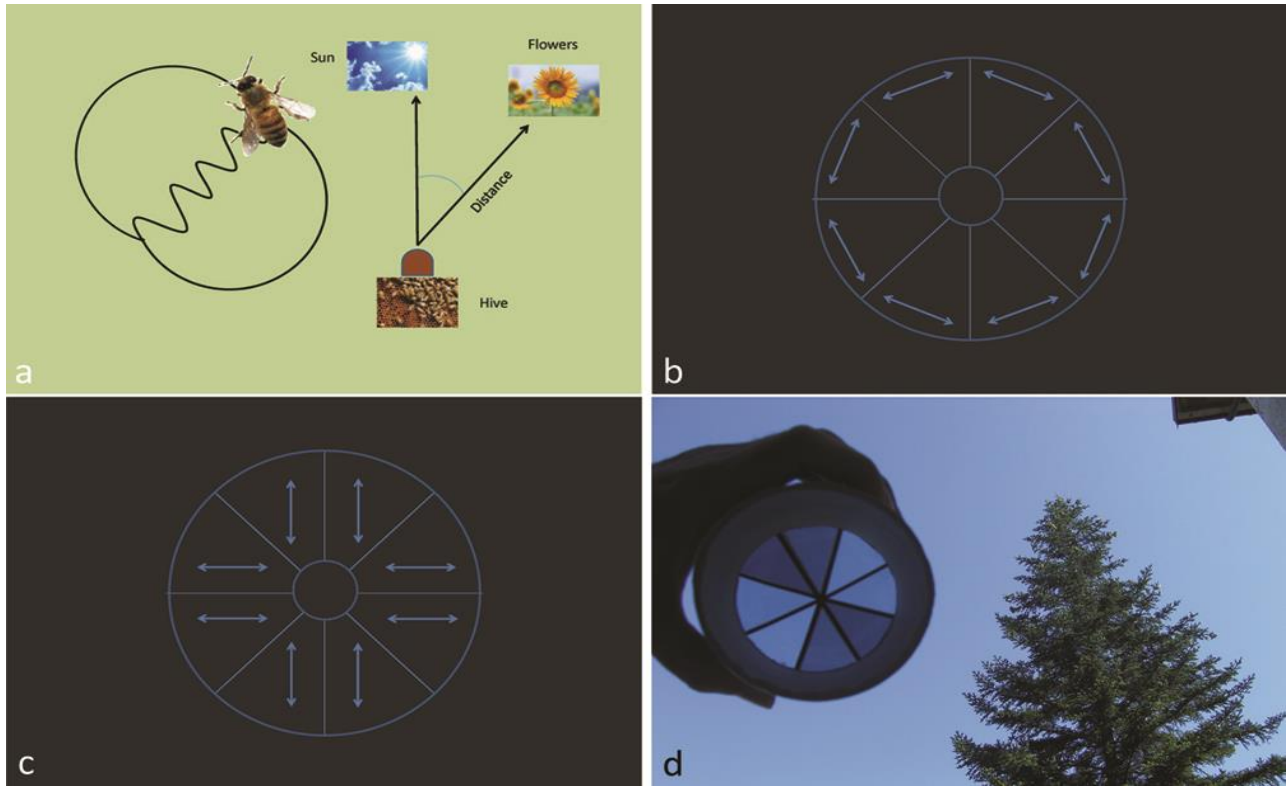

Fig. 4 (a) The waggle dance of a bee. The angle of the waggle with respect to the vertical indicates the angle made by the hive with the direction of the food and the position of the sun. (b) The model of the retinula imagined by von Frisch. The arrows indicate the sensitivity of the cells to polarization. (c) A more actual model of sensitivity (d) Von Frisch polariscope at work.

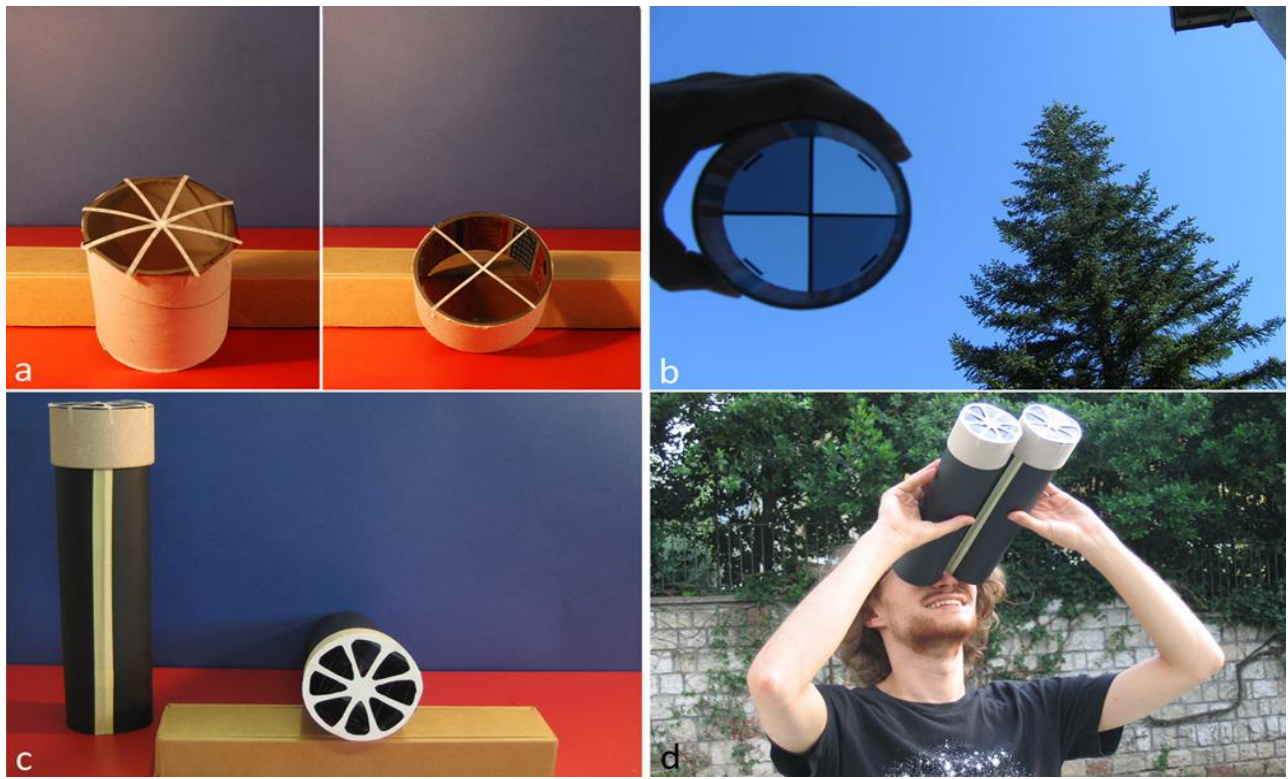

Fig. 5 (a) On the left von Frisch polariscope, and on the right our polariscope. (b) Sky polarization with our polariscope. In (c) a cardboard mask is set on our model to reproduce the disposition of the retinula cells. (d) A test of our "bipolariscope".

duration of the waggle is a measure of the distance between the food and the hive.

Honey bees, like many arthropods, have two compound eyes. They are made up of repeating units called ommatidia: in a bee's eye there are about 5000 of these structures. Each ommatidium is an independent visual system: it has a lens, a transparent crystalline cone, pigments cells which separate each ommatidia from the others, and visual cells arranged like a section of an orange forming the retinula, the corresponding structure of our retina.

According to von Frisch the eight cells of the retinula was sensitive to four directions of polarization (Fig. 4b). With this visual system bees were able to detect the position of the sun also in a cloudy day very efficiently. The Austrian scientist made also a funny model of the bee's retinula with pieces of a polarizing filter: it works quite well as a polariscope. But now we know things be a little different: in the retinula four cells are sensitive to one direction of polarization and the other four to the direction perpendicular to this one, as in Fig 4c, (and there is also a central cell) [7]. So we have made the von Frisch polariscope (Fig. 4d) but also a more realistic model of retinula with just two sensitive polarizations as in Fig 5a.

We can test our polariscope detecting the polarization of a laptop screen or the sky (Fig 5b). In Fig. 5c with a cardboard mask we have reproduced the retinula structure on our polariscope. And if you make two of them maybe you can see a little like a bee, as in Fig. 5d.

## IV. RUBBISH ART

Cellophane is made of long chains of aligned glucose molecules. Electromagnetic waves with polarization orthogonal to the molecules propagate through this material at different speed with respect to waves with polarization parallel to the chains. This is the origin of cellophane birefringence: the fact that the material has two indices of refraction depending on the light polarization. Other materials have two indices of refraction, for instance crystals with asymmetric atomic and molecular structures like calcite or tourmaline.

So in a birefringent material we can imagine a general wave as formed by two polarization components, each one depending on a different refraction index. In a calcite crystal, for instance, a incident ray of light is separate in two beams: a *slow ray* with higher refractive index and slower phase velocity, and a *fast ray* depending on the lower refractive index and having faster phase velocity. However, if the material is thin, as the cellophane, there is not a detectable separation of the first ray in two parts, the two components add up, and we can observe a rotation of the initial polarization when the resulting beam leaves the material. That is because the phase of the wave of the slow ray is retarded with respect to the other one. The polarization changes as a function of the material thickness and light wavelength.

Not only cellophane but other thin films like oriented polypropylene are birefringent, and also plastic under mechanical stress. Usually plastic is an isotropic material, but when stressed it loses its isotropy. This property is applied in mechanical engineering to detect stress in materials.

We can see beautiful pictures when we set between two crossed polarizing filters materials such as cellophane, stretched polyethylene, Plexiglas and plastic, how you can observe in our experiments represented in Fig. 6 and 7. These images remind us of those ones of abstract painting. Fig. 6a shows our simple experimental apparatus: a laptop screen is the source of polarized light [8] and several materials are set on that. A polarizing filter is placed in front of the camera.

If the polarizing filter is oriented to stop the light of the laptop, the screen looks dark. But if you set between the screen and the filter materials able to change the orientation of the laptop polarization, then it is possible to see again light, as in the Dirac experiment. If the material changes the polarization as a function of its thickness and the wavelength of light, often many beautiful colors can become visible.

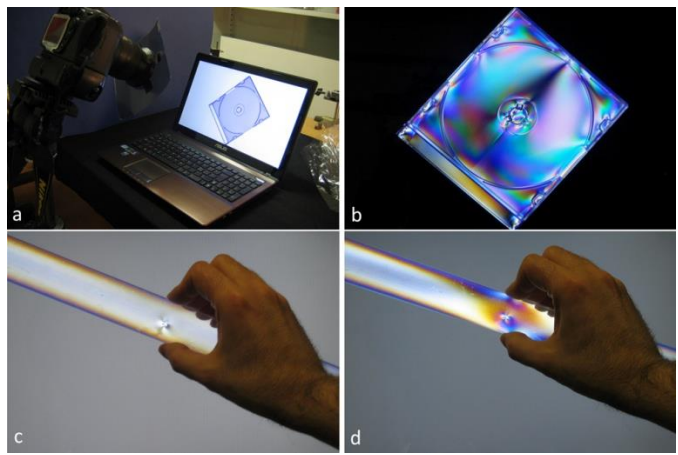

Fig. 6 (a) The apparatus: a laptop screen, a polarizing filter and a camera (b) A CD cover shows birefringence by stress. In the next pictures we have other examples of such a birefringence: in **(c)** it is showed a plastic tube without stress and in (d) the same tube with stress.

## V. SWEET EXPERIMENTS

In 1848 the French Louis Pasteur, not only a great biologist, but also a great chemist, discovered that a solution of tartaric acid rotated the light polarization. He also demonstrated the existence of molecular chirality and isomerism. Isomerism is named the existence of substance with the same chemical formula, that is, with the same atoms but with different atomic arrangement. In molecular chirality, two molecules have the same chemical formula, but they are one the mirror image of the other. Pasteur discovered that it can exist in two forms, levorotatory and dextrorotatory, by watching the two kinds of tartaric acid crystals that are one mirror image of the other. Pasteur divided the two kind of crystals and he showed that in a solution their behavior was different: they were optically active but in different way. So Pasteur imagined that also at molecular level, molecules could

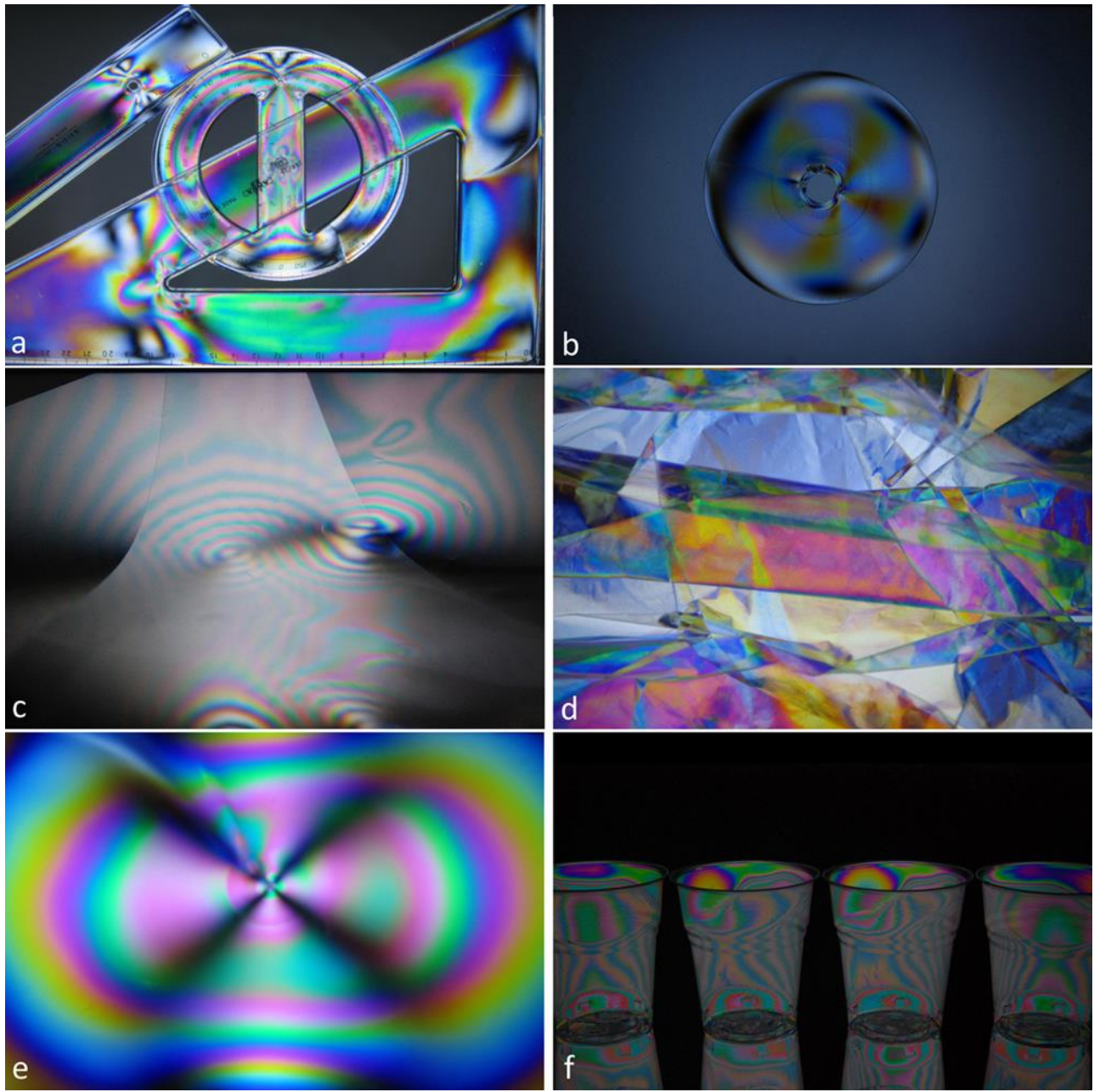

Fig. 7 (a) Birefringence at school, (b) Birefringence from a DVD protection, (c) from plastic sheets, (d) from cellophane, (e) from the bottom of a plastic box, (f) from plastic glasses.

exist with the same atoms but with mirror-like symmetry. An optically active substance rotates the plane of the light polarization (Fig. 8a). A substance is called dextrorotatory or laevorotatory if it rotates the polarization plane clockwise or anticlockwise with respect to an observer seeing the light coming. A polarized incident electromagnetic wave on an optically active substance creates an electric and magnetic dipole momentum with a different direction with respect to the incident electric field. These produce a scattered electromagnetic wave in phase with the incident wave but with different polarization. The resultant wave has the polarization rotated with respect to the incident one.

In our experiments we use cheap substances as sucrose (cane sugar) and fructose (fruit sugar). The sucrose is dextrorotatory, and the fructose is laevorotatory. Other substance that we can use is the glucose which is dextrorotatory (for this reason it is also called dextrose and fructose laevulose. The molecules of fructose and glucose are mirror images one of the other). For our experiments we use several glasses, a lamp (in Fig. 8b a slide projector), a Polaroid filter to create polarized light (but you can use also a laptop screen that usually has polarized light) and another polarizing filter to observe the effect of the optical activity. The rotation of polarization depends on the thickness of the material (in our case the size of the glass where there is the solution), and on

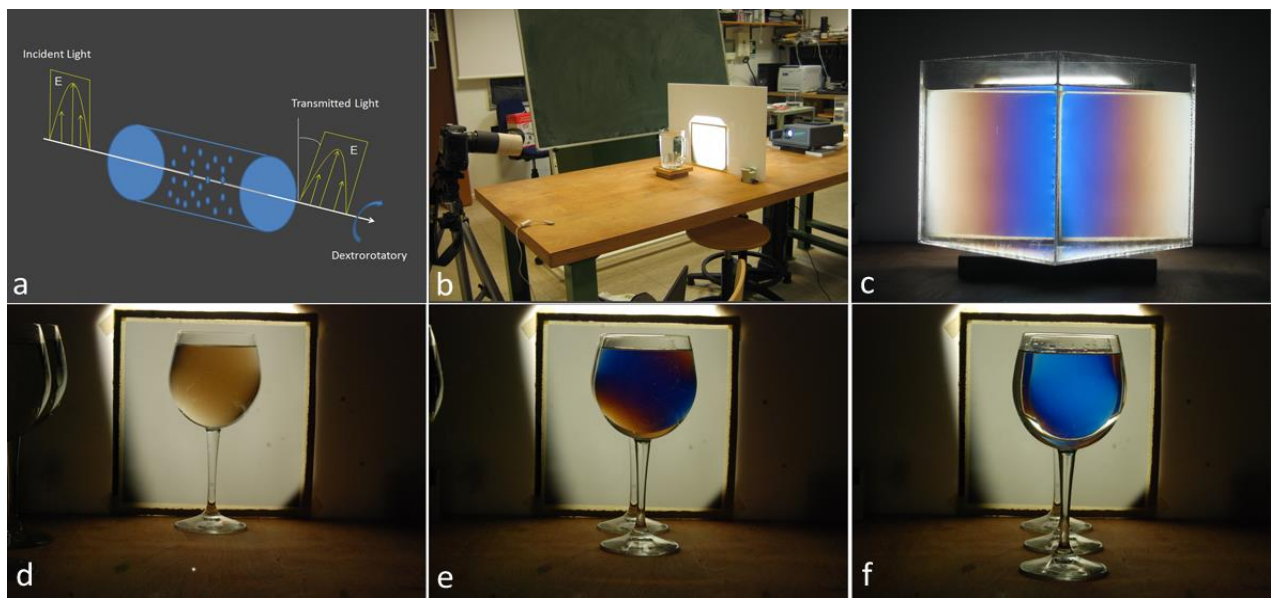

Fig. 8 (a) Schematic picture of the rotation of polarization in a dextrorotatory solution. (b) The apparatus: the polarized light comes from a slide projector having a polarizing filter in front of it. Another filter is set in front of the camera. (c) A prism full of a sugar solution. In (d)-(f) three glasses with a sugar solution are set in sequence to show how the polarization changes with the thickness of optically active material.

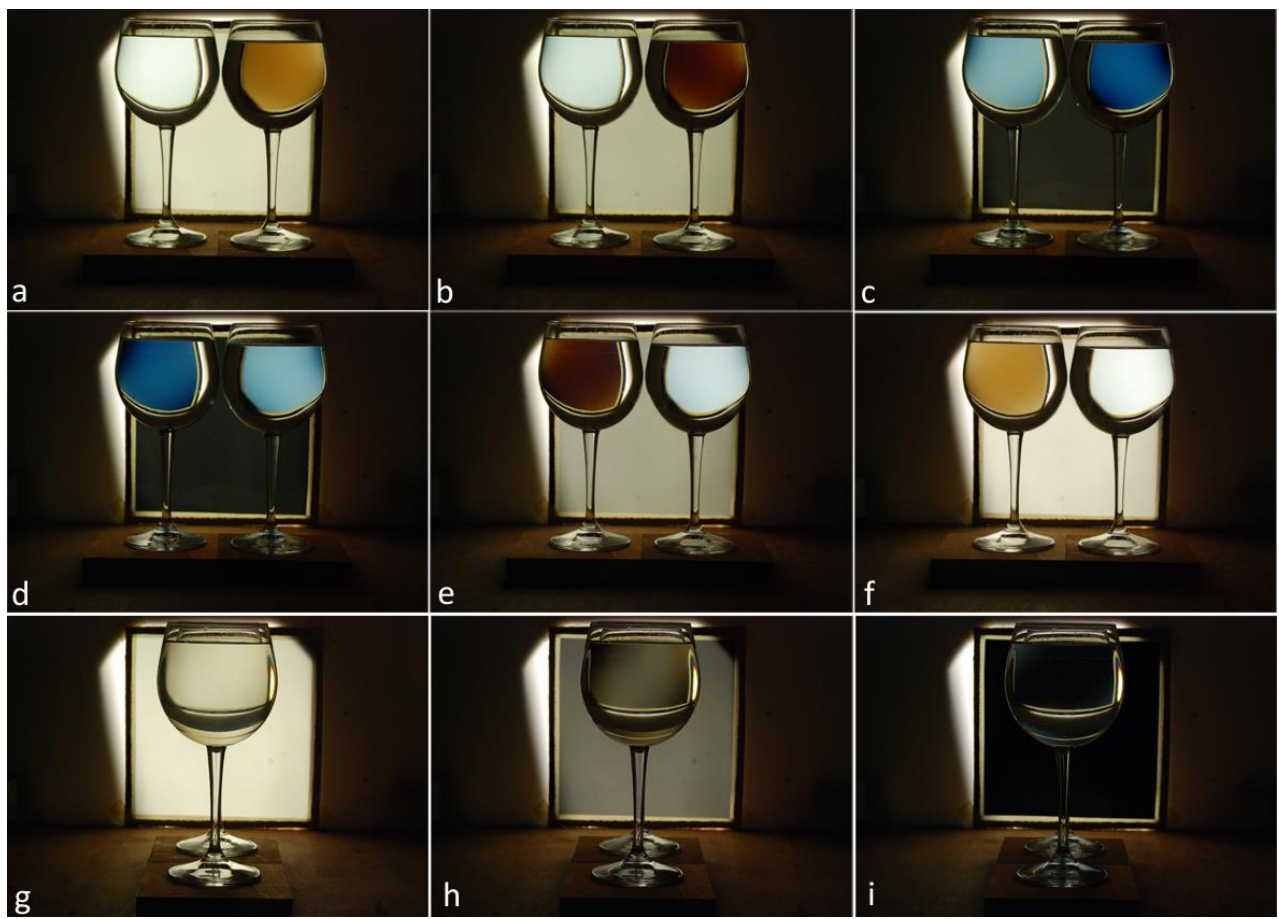

Fig. 9. (a) - (f) Rotation of polarization of fructose (on the left) and sucrose (on the right). When the Polaroid in front of the camera is rotate, we can see sequences of colours in the two solutions which are one the opposite of the other. (g) - (i) Erase effect of the two laevorotatory and dextrorotatory substances.

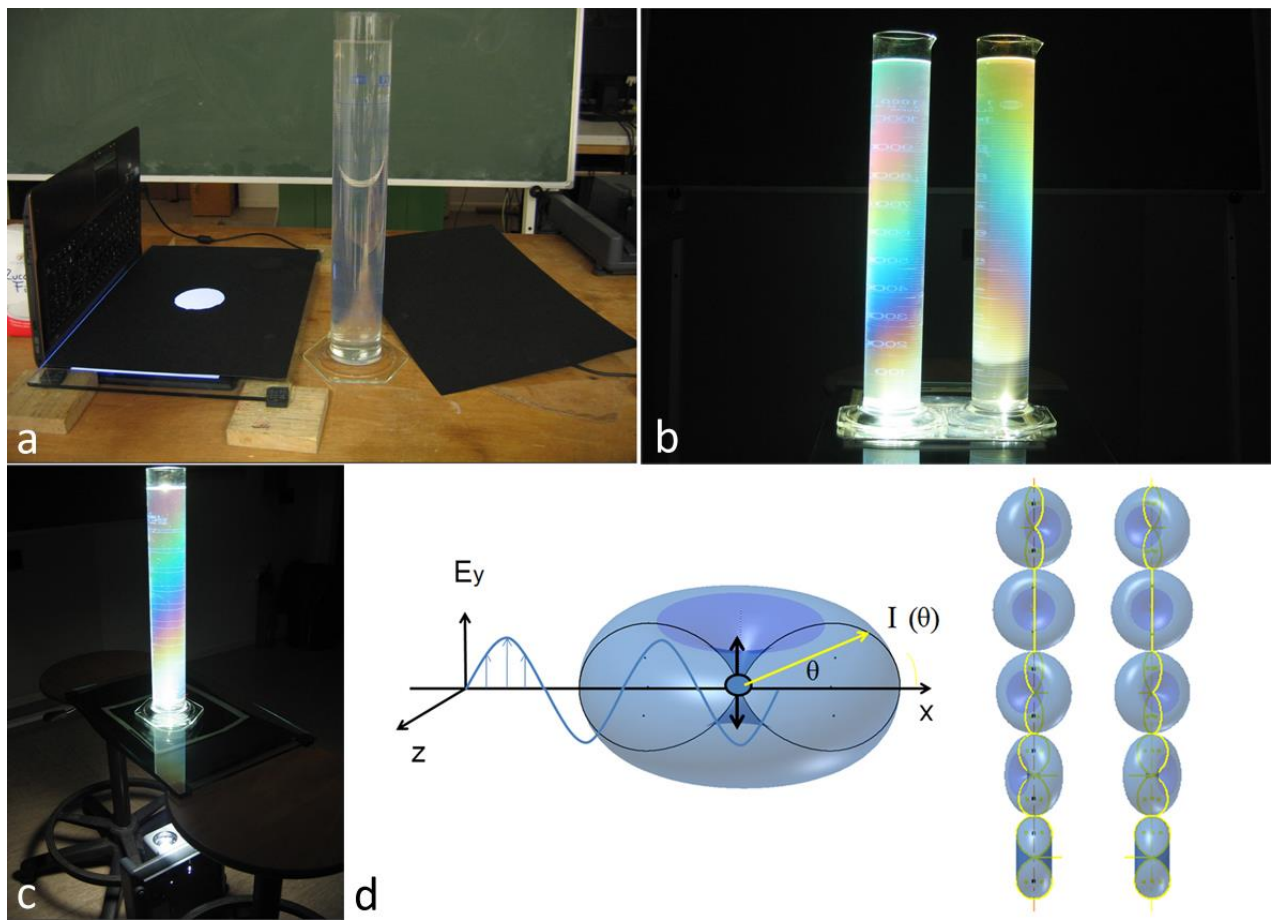

Fig. 10 (a) Experimental apparatus using the polarized light of a laptop computer. (b) Experiment with a solution of fructose (on the left) and sucrose (on the right) (c) A different configuration using the light of a slide projector and a polarizing filter. (d) Schematic distribution of the light scattered from an electric dipole and from the two columns of solutions.

the wavelength and on the concentration of the optically active substance in the solution. In Fig. 8c, for example, we have a prism with a solution of cane sugar and water. A polarizing filter is set behind that in order to create polarized light. In the picture the blue wavelength is rotate more than the red. But also we have a little component of red colour along the direction of the second Polaroid filter (set in front of the camera). You can also use a glass with the solution of cane sugar, (approximately with thickness constant, if we do not consider the effects of the borders). When the second filter is rotate, we can observe different colours, before a yellow-red and then blue. When the two filters are parallel we see again the white light. We can also check the effect of the thickness if we use three glasses, as in Fig. 8 (d)-(f). You see the yellow-red colour and then the blue.

What happens if we use a glass with a solution of fructose (with a concentration similar to the previous)? Students can verify that the colours appear with a sequence opposite to that of sucrose: the two substance rotates the polarization in the opposite sense of rotation, as the Fig. 9 (a)-(f) shows. Pupils can also experiment putting the two glasses one behind the other. They erase their effects each other and we cannot observe anymore optical activity: one glass rotates the polarization clockwise and the other anticlockwise (Fig. 9 (g) - (i)).

In order to show that the rotation of the polarization depends on the light wavelength, we can set the following experiment: we can put two glass cylinders with a solution of sucrose and fructose on a source of polarized light. What the student can see it is a helix of colours as in the Fig 10. When we change the direction of polarization the spiral moves with that. Moreover, confronting fructose with sucrose, their helixes are one the mirror image of the other (note that with the same concentration, the fructose is a little more optically active than the sucrose).

The experiment is known (see [10], but they use just one kind of solution), however we could not find an explication of it in the scientific literature. We describe the cause of the phenomenon in this way. The direction of the polarization and so the electric field is rotated by the optical activity and it makes helixes with different step with respect to the wavelength. The helix is clockwise or anticlockwise depending on the solution. The electric field excites the molecules inside the solution and those emit light. The intensity of the scattered light is zero along the direction of the incident electric field on the molecule and it has its maximum value along the directions orthogonal to that.

So the direction of the maxima of the light intensity makes a helix too. Moreover, we have different helixes for different wavelengths. When the maximum of intensity of a colour along a direction corresponds to the minima, o near to those, of others colours, we see along that direction a particular colour corresponding to a maximum of the dipole emission.

## CONCLUSIONS

Polarization has many practical applications and the treatment of this subject lends itself very well to an interdisciplinary approach, from biology to material science. Multidisciplinary can make the subject more attractive to the class, because it can also attract the attention of students not directly interested in physics. For instance, bees and other animals exploit natural polarization to sense of orientation. So teachers can talk about eye physiology, animal behavior and the polarization by scattering and reflection of light.

Moreover, it is possible to make very beautiful and intriguing experiments. How we have seen, phantasmagoric images appear when we set between two crossed polarizing filters some materials: they could be considered abstract art. Our proposal could be also a good occasion to discuss the relation between art, science and technology and how technology permits to create new forms of artistic expressions (for other newsworthy visualizations of the physics of light, you can read [11]-[13]).

### Acknowledgment

The author would like to thank for help and useful comments in the preparation of this work Mauricio Cáceres Guerrero and Antony Day of Amazon Regional University IKIAM, Ecuador, Tommaso Rosi and Luigi Gratton of Trento University and Pierpaolo Napolitano of ISTAT, Italia.